\documentclass[aps,preprintnumbers,amsmath,amssymb,endfloats]{revtex4}

\usepackage{graphicx}
\usepackage{dcolumn}
\usepackage{bm}

\begin{document}


\title{High Magnetic Field Sensor Using LaSb$_2$}

\author{D.P. Young, R.G. Goodrich, J.F. DiTusa, S. Guo, and P.W. Adams}
\affiliation{Department of Physics and Astronomy\\Louisiana State University\\Baton Rouge, Louisiana,
70803}%

\author{Julia Y. Chan}
\affiliation{Department of Chemistry\\Louisiana State University\\Baton Rouge, LA, 70803}%
\author{Donavan Hall}
\altaffiliation[Present address: ]{American Physical Society, One Research Road, Box 9000, Ridge, NY 11961}
\affiliation{National High Magnetic Field Laboratory\\Florida State University\\Tallahassee, FL,
32306}

\date{\today}

\begin{abstract} The magnetotransport properties of single crystals of the highly anisotropic layered metal
LaSb$_{2}$ are reported in magnetic fields up to 45 T with fields oriented both parallel and perpendicular to the
layers.  Below 10 K the perpendicular magnetoresistance of LaSb$_{2}$ becomes temperature independent
and is characterized by a 100-fold linear increase in resistance between 0 and 45 T with no evidence of
quantum oscillations down to 50 mK.  The Hall resistivity is hole-like and gives a high field carrier density
of n $\sim$ 3x10$^{20}$ cm$^{-3}$.  The feasibility of using LaSb$_2$ for magnetic field sensors is discussed. 
\end{abstract}

\maketitle

One of the most successful strategies for producing technologically relevant magnetoresistive materials is to
enhance the effects of field-dependent magnetic scattering processes through the creation of magnetic
superlattices \cite{GMR} or by doping magnetic insulators such that a magnetic and metal-insulator transition
coincide \cite{Manganite}.  Unexpectedly, there have been several recent discoveries of a large, non-saturating
magnetoresistance (MR) in low carrier density {\em non}-magnetic metals
\cite{Abrikosov,Lifshits,Falicov,McClure,Adams} and semiconductors \cite{Rosenbaum}.  One class of these systems, the
slightly off-stoichiometric silver chalcogenides, Ag$_{2+\delta}$Te and Ag$_{2+\delta}$Se, have shown significant
promise as the basis of ultra-high magnetic field sensors by virtue of the fact that they exhibit a multi-fold,
quasi-linear MR that remains unsaturated up to 60 T \cite{Rosenbaum}.  In this Letter we present
magntotransport data on the highly layered non-magnetic metal LaSb$_2$ which displays a 100-fold, linear MR
with no sign of saturation up to 45 T.  We show that in many respects, including sensitivity, linearity, synthesis
characteristics and intrinsic anisotropy, LaSb$_2$ is a compelling candidate for high-field sensor development.

	LaSb$_2$ is a member of the $R$Sb$_{2}$ ($R$=La-Nd, Sm) family of compounds that all form in the orthorhombic
SmSb$_{2}$ structure \cite{Wang,Hullinger}. LaSb$_2$, in particular, is comprised of alternating La/Sb layers and
two-dimensional rectangular sheets of Sb atoms stacked along the $c-$axis \cite{lattice}.  Similar structural
characteristics give rise to the anisotropic physical properties observed in all the compounds in the
$R$Sb$_{2}$ series \cite{Budko}.  Since LaSb$_{2}$ is non-magnetic, its low-temperature transport
properties are not complicated by magnetic phase transitions which occur in the other members of this series
\cite{Budko}. 

	Single crystals of LaSb$_{2}$ were grown from high purity La and Sb by the metallic flux method
\cite{Canfield}.  The orthorhombic SmSb$_{2}$-structure type was confirmed
by single crystal X-ray diffraction.  The crystals grow as large flat layered plates which are malleable and easily
cleaved.  Typically flux grown samples had dimensions of 5mm x 5mm x 0.2mm.  Electrical contact was made
using Epotek \cite{Epotek} silver epoxy and 1 mil gold wire.  Transport properties were measured using a 27 Hz
4-probe AC technique at temperatures from 0.03 - 300 K and in magnetic fields up to 45 T.  In all of the measurements
presented probe currents of 1 - 5 mA where used with corresponding power levels less than 10 nW.  Hall effect
measurements were made on natural thickness samples in a 4-wire geometry with data being taken in both positive and
negative fields up to 30 T. 

	The in-plane zero-field electrical resistivity, $\rho$, of single crystals of LaSb$_{2}$ was measured from
1.8-300 K and found to be metallic ($d\rho/dT>0$).  The residual resistivity ratio (RRR) was large
($\rho_{ab}(300 K)/\rho_{ab}(2 K)\approx70-90$), indicating a high sample quality.  In the main panel of Fig.\ 1
we show the transverse MR with the field oriented parallel and perpendicular to the $ab$-plane.  Both MR's
are positive and nearly linear above 2 T.  Note the extreme anisotropy in the magnetotransport with the
perpendicular MR being an order of magnitude larger than the parallel MR.  The perpendicular MR was large,
with resistance increasing by a factor of 90 from 0 to 45 T.  The MR was temperature independent below 10 K but
decreased rapidly above 30 K.  Interestingly, there is no evidence of saturation or quantum oscillations in the MR
of Fig. 1 which has obvious advantages for magnetic field sensor applications. The solid lines in Fig. 1 are
least-square fits to a fourth-order polynomial (Table 1). Note the high quality of the fits to the MR using this
simple functional form.  The relative field sensitivity, which is the figure of merit for a sensor, is represnted by
$\alpha_1=1.23$ $\rm T^{-1}$.  This sensitivity is roughly a factor of 4 greater than that of
Ag$_{2+\delta}$Se (triangle symbols in Fig.\ 1).

	The anisotropy of the MR can be demonstrated by measuring the transverse MR as a function of the
tilt angle in a constant magnetic field.  In Fig.\ 2 we plot the resistivity normalized by
the parallel field value, $\rho(H_{\perp}=0)$, as a function of the perpendicular component of the field
$H_{\perp}$.  Interestingly, the general shape of the MR curves in Fig.\ 2 is quite similar to those of Fig.\
1.  The solid lines are polynomial fits to the data (Table 1).  In terms of a magnetic field sensor, this
anisotropy can be exploited to determine the angle of orientation in tilted field studies.  The micaceous nature of
LaSb$_2$ not only produces an anisotropic MR but also presents a convenient geometry for Hall measurements, namely
large flat crystals.  In the main panel of Fig.\ 3 we plot the Hall resistivity, $\rho_{xy}$, as a function of
magnetic field at T = 2 K.  In the right inset of Fig.\ 3 we show the low field behavior of $\rho_{xy}$ which is
negative below 0.5 T but becomes positive at higher fields.  This latter behavior is often characteristic of a
two-carrier system \cite{Ashcroft}.  In the high field limit the majority carrier dominates, which in our case is
hole-like.  Above 10 T the Hall constant is R$_H$ $\sim$ 2x10$^{-6}$ $\mu\Omega$-cm/T which corresponds to a carrier
density of n $\sim $3x10$^{20}$ cm$^{-3}$ and a Hall mobility $\mu\sim 0.05$ m$^2$/Vs.  We note that the overall
shape of the Hall resistance curve in the right inset of Fig.\ 3, with its local minimum, is very
similar in character to that of NbSe$_2$ and TaSe$_2$ which also form in non-magnetic micaceous crystalline
structures \cite{Naito}.  

	A useful measure of the magnitude of the linear MR is the dimensionless Kohler slope,
$S=(1/R_H)(d\rho(H)/dH)$.  Combining the Hall constant measurements with the value of $\alpha_1$, obtained from the
MR of Fig.\ 1, we get a high field value S $\sim$ 0.6.  This value is an order of magnitude larger than what is
typical of other non-magnetic systems displaying a linear MR.  Classically, the MR should vary
quadratically with field.  In a closed orbit system the MR saturates in the high field limit limit,
$\omega_c\tau\gg 1$, where $\omega_c$ is the cyclotron frequency and
$\tau$ is the elastic scattering time \cite{Ashcroft}.  Over the past 30 years several mechanisms have been proposed
to account for anomalous linear MR observed in a wide variety non-magnetic systems such as elemental metals
\cite{Kapitza,Falicov}, two-dimensional heterostructures \cite{Stormer}, and disordered semiconductors
\cite{Rosenbaum}.  Theories accounting for linear MR fall into two main categories.  The first contains
theories associated with the alteration of the structural symmetry due to the formation of a charge density wave
(CDW) \cite{Overhauser}.   Linear MR in very pure elemental metals has been attributed to quantum fluctuations about
a CDW ground state \cite{Neto} and/or a magnetic breakdown of the CDW gap \cite{Wilson,Naito}.  The second includes
theories which invoke high field quantization effects or singular scattering mechanisms which cannot be accounted
for by the standard perturbative scattering formulations
\cite{Abrikosov,Young}.  Interestingly, the transition metal dichalogenides NbSe$_2$ and
TaSe$_2$ both have well established CDW ground states and exhibit an anomalous linear MR.  These compounds are
similar in structure to LaSb$_2$, suggesting that perhaps a CDW state plays a central role in the MR of LaSb$_2$.  At
this time, however, is not known whether LaSb$_2$ undergoes a charged density wave transition.

	It has been known for many years that the relative MR, $\Delta\rho/\rho$, of many metals and semimetals is a
temperature independent function of magnetic field \cite{Kohler}.  In particular, $\Delta\rho/\rho=F(H)$, where
$F(H)$ usually has a power-law form.  LaSb$_2$ is known to obey this rule, commonly refered to as Kohler's rule, with
$F(H)\sim H$
\cite{Budko}.  One can also make a similar analysis by substituting the the Hall resistance for the magnetic field
$H$.  The resulting modified Kohler plot for LaSb$_2$ is shown in the left inset of Fig.\ 3.  The solid line in the
plot has a slope of $\nu=2/3$ indicating that $\Delta\rho\propto{\rho_{xy}}^{2/3}$.  Interesingly, Ag$_{2+\delta}$Se
also exhibits power-law behavior but with a low temperature modified Kohler slope of $\nu=5/3$ \cite{Rosenbaum}. 
This suggests that the underlying MR mechanisms in these two seemingly unrelated systems may be similar and that
the differing scaling exponents is a dimensionality effect.   

	In conclusion, we show that crystalline LaSb$_2$ is an attractive material for use as a high magnetic field
sensor.  It is relatively easy to synthesize, and electrical contact can be made with silver epoxy.  By virtue of
its highly anisotropic structure, LaSb$_2$ can be used in either a transverse MR configuration or a Hall
configuration.  The sensitivity in both configurations is quite good.  Calibration curves for both the MR and the
Hall resistance can be made using a fourth-order polynomial, thus avoiding numerical difficulties associated with
complicated fitting forms.  

	We gratefully acknowledge discussions with Dana Browne and Richard Kurtz. This work was supported by
the National Science Foundation under Grants DMR 99-72151 and DMR 01-03892.  We also acknowledge support of the
Louisiana Education Quality Support Fund under Grant No. 2001-04-RD-A-11.  A portion of this work was performed at
the National High Magnetic Field Laboratory, which is supported by NSF Cooperative Agreement No. DMR-9527035 and
by the State of Florida.

\newpage
\begin{figure}
\includegraphics[width=5in]{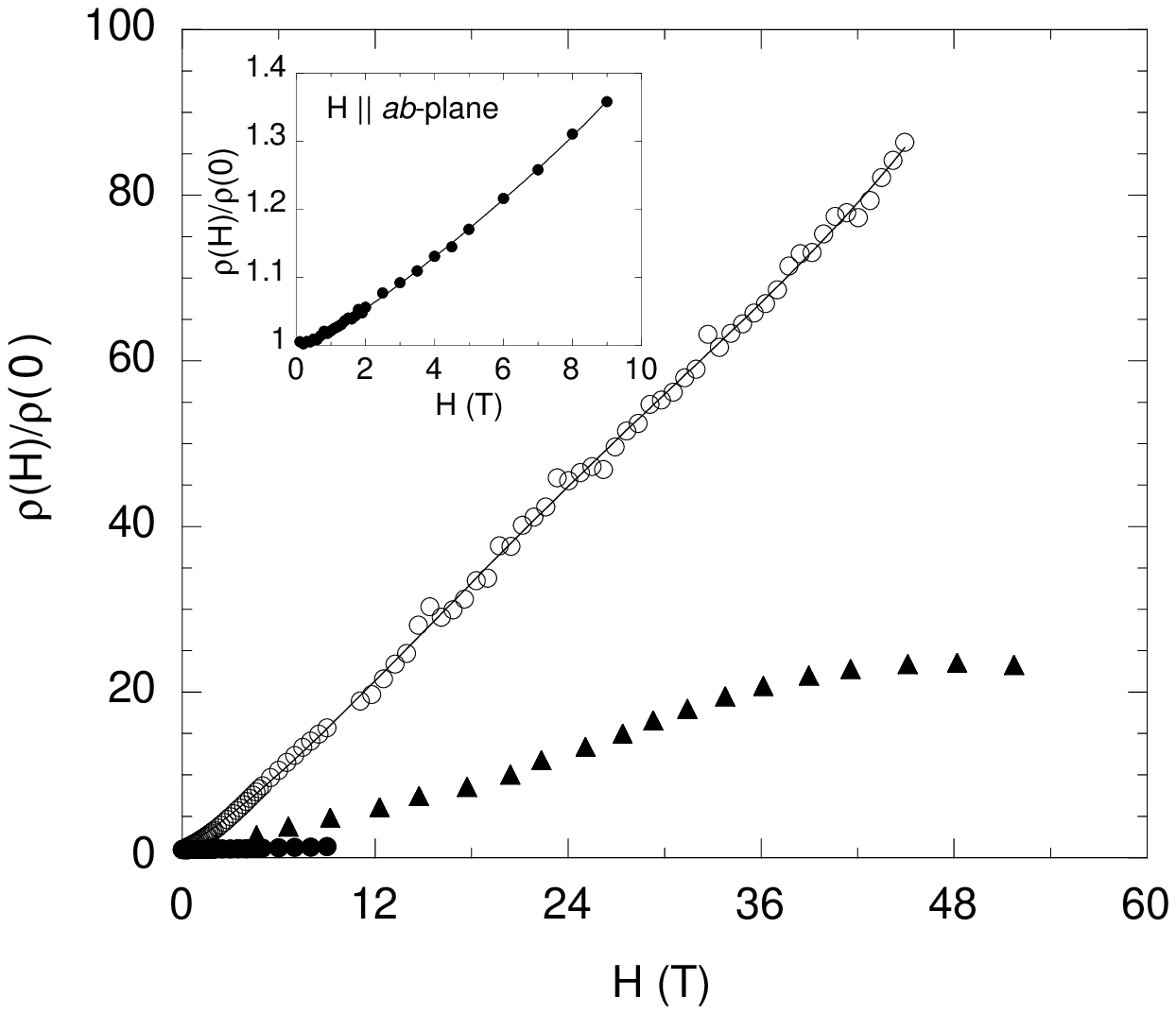}
\caption{\label{fig:epsart} Transverse MR of LaSb$_2$ at T = 2 K with the current in the $ab$-plane and magnetic
field oriented parallel (closed circles) and perpendicular (open circles) to the $ab$-plane.  The solid triangles
represent the MR of Ag$_{2+\delta}$Se as taken from Ref.\ 8.  Inset: Low field MR with
$H\| ab$-plane.  The solid lines represent a least-squares fit to the data using a fourth-order polynomial (Table
1).
}
\newpage
\end{figure}

\begin{figure}
\includegraphics[width=5in]{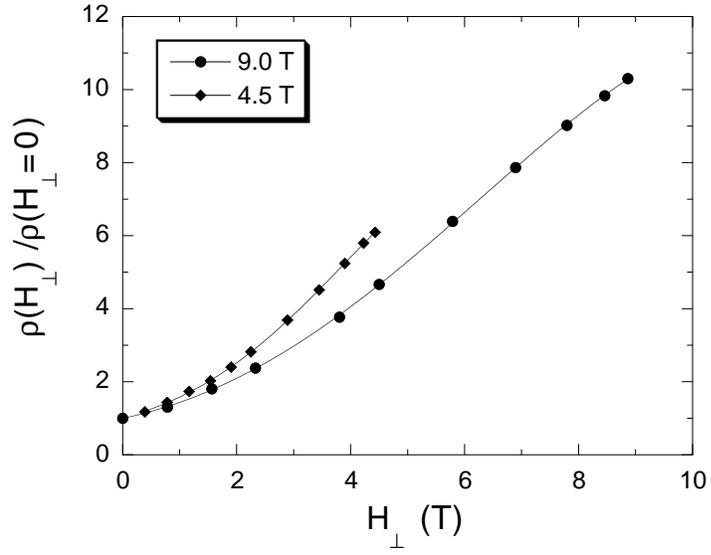}
\caption{\label{fig:epsart} MR in a tilted magnetic field. The samples were rotated out of the $H\parallel ab$-plane
in  constant magnetic fields of 9.0 T and 4.5 T.  The MR at 2 K is plotted as a function of the perpendicular
component of the field.  The solid lines represent a least-squares fit to the data using a fourth-order polynomial
(Table 1).
}
\newpage
\end{figure}

\begin{figure}
\includegraphics[width=5in]{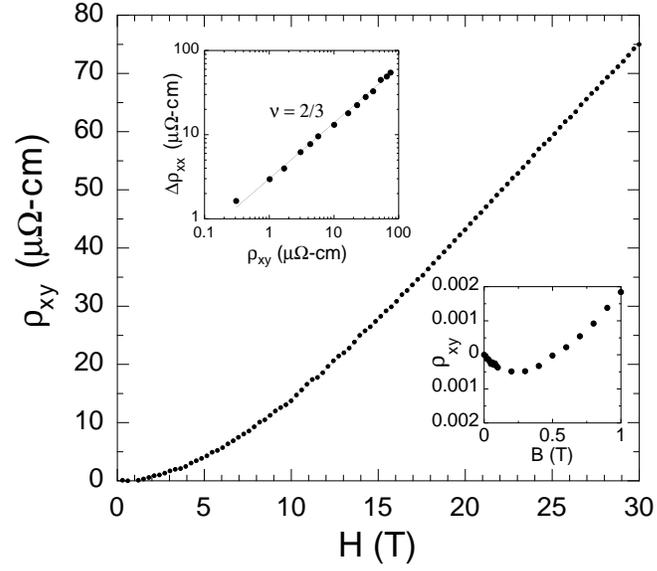}
\caption{\label{fig:epsart} Hall resistivity as a function of magnetic field at T = 2 K. The solid lines represent a
least-squares fit to the data using a fourth-order polynomial (Table 1).  Right inset: low field behavior of
$\rho_{xy}$ showing a crossover from electron-like to hole-like conduction with increasing field.  Left inset:
log-log plot of $\Delta\rho = \rho(H)-\rho(0)$ as a function of the Hall resistivity at T = 2 K. The solid line has
slope $\nu=2/3$. The linearity of the data suggests that $\Delta\rho\propto{\rho_{xy}}^{2/3}$.
}
\newpage
\end{figure}

\begin{table}
\caption{\label{tab:table1} Polynomial coefficients obtained from a least squares fit to the data in Figs.\
1-3 using a fourth-order polynomial, $f(H)=\alpha_0+\alpha_1H+\alpha_2H^2+\alpha_3H^2+\alpha_4H^4$.}
\begin{ruledtabular}
\begin{tabular}{cccccc}
 &$\alpha_0$&$\alpha_1$ &$\alpha_2$ &$\alpha_3$&$\alpha_4$ \\
\hline
$\rho(H)/\rho(0)$ $(H\parallel\hat{c})$ & 1& 1.236 & 0.0579 & -1.874x10$^{-3}$ & 2.019x10$^{-5}$ \\
$\rho(H)/\rho(0)$ $(H\perp\hat{c})$ & 1& 0.0182 & 5.601x10$^{-3}$ & -6.260x10$^{-4}$ & 3.032x10$^{-5}$ \\
$\rho(H_{\perp})/\rho(0)$ $(H=9\rm T)$ & 1 & 0.3315 & 0.1023 & 4.598x10$^{-3}$ & -7.931x10$^{-4}$ \\
$\rho(H_{\perp})/\rho(0)$ $(H=4.5\rm T)$ & 1& 0.4624 & 0.0827 & 0.04644 & -6.775x10$^{-3}$ \\
$\rho_{xy}(H)$ $(\mu\Omega$-cm$)$ &0 & 0.0225 & 0.1781 & -4.309x10$^{-3}$ & 3.757x10$^{-5}$ \\
\end{tabular}
\end{ruledtabular}
\end{table}


\begin{references}
\bibitem{lattice}X-ray analysis determined the lattice constants to be 0.6319 nm, 0.6173 nm,
and 1.857 nm for the $a$, $b$, and $c$ crystallographic directions, respectively.
\end{references}
\end{document}